**Terahertz magneto-photocurrents in the topological insulator Bi$_2$Se$_3$ probe its topological surface states**

Chihun In,[1,2,*] Genaro Bierhance,[1,2] Deepti Jain,[3] Tom S. Seifert,[1] Oliver Gueckstock,[1] Roberto Mantovan,[4] Seongshik Oh,[3,5] Tobias Kampfrath[1,2]

[1]Department of Physics and Halle-Berlin-Regensburg Cluster of Excellence CCE, Freie Universität Berlin, 14195 Berlin, Germany

[2]Department of Physical Chemistry, Fritz Haber Institute of the Max Planck Society, 14195 Berlin, Germany

[3]Department of Physics and Astronomy, Rutgers, The State University of New Jersey, Piscataway, NJ 08854, USA

[4]CNR-IMM, Unit of Agrate Brianza, Agrate Brianza 20864, Italy

[5]Center for Quantum Materials Synthesis, Rutgers, The State University of New Jersey, Piscataway, NJ 08854, USA

*inc@zedat.fu-berlin.de

**Abstract.** We study ultrafast magneto-photocurrents in a three-dimensional topological insulator. For this purpose, we excite (In$_r$Bi$_{1-r}$)$_2$Se$_3$ thin films with a femtosecond laser pulse in the presence of an external magnetic field $\boldsymbol{B}_\mathrm{ext}$ up to 0.3 T parallel to the film plane. The resulting in-plane photocurrent is measured by detecting the emitted terahertz (THz) electromagnetic pulse. It scales linearly with $\boldsymbol{B}_\mathrm{ext}$ and is perpendicular to $\boldsymbol{B}_\mathrm{ext}$. Strikingly, for $r \geq 4\%$, we observe an abrupt photocurrent reduction, which is strongly correlated with the Indium-induced quenching of the topological surface states. The rise time, decay time and amplitude of the THz magneto-photocurrent can consistently be explained by a scenario in which optically excited spin-polarized electrons propagate toward the film surface where the accumulated spin is converted into an in-plane charge current due to spin-velocity locking. Our results are highly relevant for contact-free probing of spin-charge conversion in systems with paramagnetic rather than spontaneous magnetic order.

Three-dimensional topological insulators (TIs) like $Bi_2Se_3$, $Bi_2Te_3$ and $Sb_2Te_3$ are fascinating materials for both fundamental and applied reasons [1–5]. While the bulk of an ideal TI is insulating, the surface hosts topologically protected electronic surface states (TSS) with high conductivity and spin-velocity locking. As TSS enable efficient spin-to-charge-current conversion (SCC) [6–11], TIs are highly interesting for spintronic applications. Examples include magnetic-order switching by electric currents [12–15] and the generation of terahertz (THz) electromagnetic pulses from femtosecond spin transport [16–19].

However, in many of these studies [8,12,13,16], the TI was in contact with a ferromagnet or ferrimagnet whose impact on the TI surface states has remained elusive [20]. Therefore, methodologies to probe SCC in TIs without additional perturbing layers are highly desirable.

**In this work**, we use femtosecond optical pulses to drive photocurrents in thin films of the model TI $Bi_2Se_3$ under the influence of an external static in-plane magnetic field $\boldsymbol{B}_\text{ext}$ [Fig. 1(a)]. By detecting the resulting THz emission, we find in-plane-photocurrents with an amplitude proportional to $\boldsymbol{B}_\text{ext}$ and a direction perpendicular to $\boldsymbol{B}_\text{ext}$. Strikingly, as 20% of Bismuth is substituted for Indium and the TSS are quenched, the amplitude of the magneto-photocurrents drops by more than 95%.

We can consistently explain our observations by a scenario in which the optical pulse launches an out-of-plane electric current that has a spin polarization of $\sim 0.3\%$ owing to the Pauli-type paramagnetism of $Bi_2Se_3$ [21,22]. As the current arrives at the surface, it is converted into an in-plane charge current. The observed charge current is sizeable because of the large SCC in $Bi_2Se_3$ [16–19]. Our results are highly interesting for contact-free probing of SCC in materials and sample systems with paramagnetic rather than spontaneous magnetic ordering.

**Samples.** Our $Bi_2Se_3$ films (thickness $d = 10\text{-}50$ QL with 1 QL = 1 quintuple layer $\approx 1$ nm) are grown by molecular-beam epitaxy on $Al_2O_3$ substrates (thickness 0.5 mm) and capped by a Selenium layer (20 nm) for protection. We also study $(In_rBi_{1-r})_2Se_3$(30 QL) thin films whose TSS are known to disappear for an Indium fraction of $r > 7\%$ [23–25].

Identically grown samples were characterized extensively in previous works [23,26,27]. In particular, angle-resolved photoemission spectroscopy (ARPES) [28] revealed that the Fermi energy $\epsilon_\text{F}$ is located 15 meV above the bottom of the bulk conduction band, corresponding to about 0.4 eV above the TSS Dirac point. Our THz Hall measurements (Supplemental Section I) are consistent with previous DC results [23,26] and show an impact of $r$ on the conduction-electron density.

**THz setup.** In our setup, an external magnetic field $\boldsymbol{B}_\text{ext} = B_\text{ext}\boldsymbol{u}_y$ with magnitude up to 0.3 T is applied along the $y$ axis with unit vector $\boldsymbol{u}_y$ parallel to the TI film plane [Fig. 1(a)]. The sample is excited by femtosecond laser pulses normally incident from the substrate side, thereby launching a photocurrent with density $\boldsymbol{j}(z,t)$ in the TI. The current gives rise to the emission of an electromagnetic pulse with frequencies extending into the THz range. The THz electric field is characterized by electro-optic sampling in a 1-mm-thick ZnTe(110) window. The THz field component along $\boldsymbol{u}_x$ and $\boldsymbol{u}_y$, respectively, results in an electro-optic signal $S_x(t, B_\text{ext})$ and $S_y(t, B_\text{ext})$. All measurements (Figs. 1-3, Supplemental Sections II-IV) are conducted under ambient conditions in a dry-air atmosphere.

**Photocurrent and contributions.** The THz signal $S_x$ can be used to retrieve the current $J_x(t) = \boldsymbol{J}(t) \cdot \boldsymbol{u}_x$ of the photocurrent sheet density $\boldsymbol{J}(t) = \int \text{d}z\, \boldsymbol{j}(z,t)$ flowing inside the sample [Fig. 1(a)] [29].

Note that a nonzero $\boldsymbol{J}(t)$ requires inversion asymmetry (IA) of the sample. Here, we focus on in-plane magneto-photocurrents due to sample-intrinsic IA (SIA). To minimize contributions by light-induced IA

(LIA), in particular, due to pump-intensity gradients inside the sample, we use films thinner than 50 nm. Data from thicker samples are addressed in Supplemental Section IV.

To minimize signals from (i) $z$-directed photocurrents and (ii) in-plane shift currents, we (i) choose nearly normal incidence of the pump and (ii) set one of the mirror axes of the $Bi_2Se_3(111)$ plane parallel to $\boldsymbol{u}_y$ [Fig. S3(a)] [30].

**Typical THz emission data.** Figure 1(b) shows typical THz waveforms $S_x(t, B_{ext})$ from $Bi_2Se_3$(16 QL) for opposite directions of the external magnetic field of $B_{ext} = \pm 0.3$ T. Remarkably, the signal partially reverses when $B_{ext}$ is reversed.

Formally, one can write the THz signal $S_x(t, \pm B_{ext}) = S_{x+}(t) \pm S_{x-}(t)$ in terms of even ($S_{x+}$) and odd ($S_{x-}$) components under reversal of $B_{ext}$. In other words,

$$S_{x\pm}(t) = \frac{S_x(t, +B_{ext}) \pm S_x(t, -B_{ext})}{2}. \quad (1)$$

In the following, we focus on the odd-in-$B_{ext}$ part $S_{x-}(t)$ [Fig. 1(c)] because it arises from a magnetic-field-dependent photocurrent, i.e., a magneto-photocurrent.

We find that (i) $S_{x-}$ along $\boldsymbol{u}_x$ is about 50 times larger than the THz signal $S_{y-}$ along $\boldsymbol{u}_y$ [Figs. 1(c) and S4(b)]. Therefore, the associated photocurrent $\boldsymbol{J}(t)$ is perpendicular to the external magnetic field $\boldsymbol{B}_{ext} = B_{ext}\boldsymbol{u}_y$. (ii) $S_{x-}$ reverses nearly completely when the sample is turned by 180° [Figs. 1(c) and S7] and, therefore, predominantly arises from SIA. Further, (iii) $S_{x-}$ is proportional to $B_{ext}$ [Figs. 1(d), S2(d) and S2(f)] and the pump power [Fig. S2(h)], but it is (iv) independent of the $Bi_2Se_3$ azimuthal angle [Fig. S3(c)] and the pump polarization [Fig. S3(d)]. (vi) The presence of the Se layer results in only minor changes to $S_{x-}(t)$ (Fig. S5).

(vii) The even-in-$B_{ext}$ part $S_{x+}(t)$ is independent of $B_{ext}$ and, thus, equals $S_x(t)$ for $B_{ext} = 0$ T. In other words, $S_{x-}$ and $S_{x+}$ have very different origin. In fact, $S_{x+}$ predominantly arises from a shift current [30], and its sizable amplitude still remains because the $Bi_2Se_3$ mirror axis is not perfectly aligned along $\boldsymbol{u}_y$ (Supplemental Section II-A).

(viii) Any $z$-directed photocurrent $J_z(t)$ is detected by tilting the sample, such that $J_z(t)$ is partially projected onto the plane of the electro-optic detection [30]. Because $J_z(t)$ only very weakly changes with $B_{ext}$, it is indistinguishable from the shift current (Supplemental Section VI).

**Impact of $r$.** To gain more insights into the origin of the magneto-photocurrent $J_{x-}(t)$ that drives $S_{x-}(t)$, we measure $S_{x-}(t)$ from $(In_rBi_{1-r})_2Se_3$(30 QL) [Fig. 2(a)-(e)] for $B_{ext} = \pm 0.3$ T. In these films, the substitution of Bismuth by Indium modifies the TSS by replacing Bismuth p-orbitals (6p$^3$) with Indium s- and p-orbitals (5s$^2$5p$^1$) and by weakening the spin-orbit coupling [23]. Indeed, previous ARPES measurements on samples from the same batch [25] confirm that the TSS bandgap opens for $r > 7\%$ [Fig. 2(f)-(j)].

Figs. 2(a)-(e) reveal that the THz emission signals $S_{x-}(t)$ have approximately identical shape for all $r$ [Fig. S6(c)]. In contrast, their amplitude exhibits a drastic reduction to 71%, 42% and 4%, respectively, as $r$ increases from 7% to 10% to 20%. Strikingly, the monotonic amplitude reduction of $S_{x-}(t)$ [Fig. 2(a)-(e)] is strongly correlated with the decreasing intensity of the TSS in the ARPES spectra [Fig. 2(f)-(j)].

**Impact of $d$.** Finally, we vary the $Bi_2Se_3$ thickness $d$ from 10 QL to 50 QL. The shape of the THz signal waveforms $S_{x-}(t)$ changes significantly with $d$ [Fig. 3(a)]. At the same time, the sign of the THz signal reverses when the sample is turned by 180° about $y$-axis (Fig. S7). Figure 3(b) displays the signal amplitude that exclusively arises from the sample's SIA (Supplemental Section IV). The amplitude increases up to $d = 30$ QL, but, importantly, saturates at larger thicknesses.

It is instructive to extract the dynamics of the charge-current sheet density $J_{x-}(t)$ from $S_{x-}(t)$ [Fig. 3(c)] [29]. Remarkably, as $d$ increases from 10 QL to 30 QL and 50 QL, the rise and decay of $J_{x-}(t)$ slow down significantly.

**Interpretation.** Our THz-emission data imply five essential findings: (F1) The magneto-photocurrent $J_{x-}\boldsymbol{u}_x$ is perpendicular to $\boldsymbol{B}_{\text{ext}} = B_{\text{ext}}\boldsymbol{u}_y$ and grows linearly with $B_{\text{ext}}$ [Fig. 1(d)]. (F2) $J_{x-}$ drops abruptly as the Indium fraction in (In$_r$Bi$_{1-r}$)$_2$Se$_3$ rises [Fig. 2(a)-(e)] and the TSS intensity in the ARPES spectra is quenched [Fig. 2(f)-(j)]. (F3) $J_{x-}$ is independent of the direction of the linear pump polarization [Fig. S3(d)]. (F4) The THz signal sign changes when the sample is turned by 180° [Figs. 1(c) and S7]. (F5) The relaxation time of $J_{x-}$ increases with the sample thickness.

Finding (F1) suggests that the magneto-photocurrent arises from Hall-effect-type physics, i.e., a primary current $j_z$ along the $z$ axis, followed by its conversion into an $x$-directed charge current $j_x$ [Fig. 4(a)]. Finding (F2) suggests that this conversion is related to TSS and, thus, SCC. It is not related to the changed pump conditions because the pump absorptance of (In$_r$Bi$_{1-r}$)$_2$Se$_3$(30 QL) remains between 37% and 41% for $r \geq 7\%$ [Fig. S6(a)]. Finding (F3) shows that shift currents make a minor contribution to $J_{x-}$. Finding (F4) implies that the signal arises predominantly from the SIA of the unexcited sample rather than from pump-intensity gradients along the $z$ axis, which may, e.g., induce a temperature gradient and, thus, a Seebeck-like current [31]. Finding (F5) suggests that the signal involves charge carriers throughout the TI thickness.

To explain our findings, we consider two scenarios (A) and (B). Briefly, both (A) and (B) rely on a bulk electron current $j_z$ of photoexcited electrons that is driven by the out-of-plane space-charge field of our Bi$_2$Se$_3$ film [Fig. 4(a)]. In (A), $j_z$ is spin-polarized by $B_{\text{ext}}$. Its SCC [32] at the Bi$_2$Se$_3$ surface results in a surface current $j_x$. In (B), the ordinary Hall effect (ordHE) converts $j_z$ into a current $j_x \propto B_{\text{ext}}$ in the bulk [33]. While both scenarios are consistent with the symmetry properties of signal $S_{x-}(t)$, scenario (A) explains quantitative features of our findings significantly better. We first consider scenario (A) and subsequently discuss possible contributions of mechanism (B).

**Scenario (A).** The idea is outlined in Fig. 4(a): (A1) The external magnetic field $\boldsymbol{B}_{\text{ext}}$ causes a small but nonzero spin polarization $P_s \propto B_{\text{ext}}$ of the Bi$_2$Se$_3$ bulk electrons. (A2) Photoexcitation together with the built-in space-charge field of the Bi$_2$Se$_3$ film launches an electron current $j_z\boldsymbol{u}_z$ with spin polarization $P_s$ toward the Bi$_2$Se$_3$/air interface. (A3) The associated spin-current density $j_s = j_zP_s$ induces a spin-polarized occupation of the TSS. Owing to spin-velocity locking, a charge current with sheet density $J_{x-}$ perpendicular to $B_{\text{ext}}$ results. Indeed, TSS are known to have a large SCC efficiency [34]. In the following, we discuss the plausibility of steps (A1)-(A3) in the framework of a simple analytical model.

(A1) We assume Pauli paramagnetism where the external magnetic field shifts spin-up and spin-down bands of the Bi$_2$Se$_3$ bulk by an energy $\pm g_s\mu_B B_{\text{ext}}$. Here, $g_s \approx 2$ is the g-factor of the free electron, and $\mu_B$ is the Bohr magneton. The induced spin density $N_\uparrow - N_\downarrow = g_s\mu_B B_{\text{ext}} D_e(\epsilon_F)$ scales with the electronic density of states $D_e$ at the Fermi energy $\epsilon_F$. In our sample, the known band structure [35,36] and known $\epsilon_F \sim 15$ meV [26] imply a spin polarization of $P_s = (N_\uparrow - N_\downarrow)/(N_\uparrow + N_\downarrow) \approx 0.3\%$ for $B_{\text{ext}} = 0.3$ T. The initial pump-induced photoexcited spin density of $\Delta(N_\uparrow - N_\downarrow)_0 \sim 10^{22}$ m$^{-3}$ is estimated by multiplying the excited electron density $\Delta(N_\uparrow + N_\downarrow)_0$ with $P_s$.

(A2) Bi$_2$Se$_3$ has an internal space-charge field $\boldsymbol{E}_{\text{SC}} = E_{\text{SC}} \boldsymbol{u}_z$ along the $z$ axis. It arises from positively charged Selenium vacancies at the Bi$_2$Se$_3$ interface and can be reasonably well modeled by $E_{\text{SC}}(z) = -Z_0 c e n_0 z / \varepsilon_{\text{DC}}$ [see Figs. 4(b) and S10] [26,37,38]. Here, $n_0$ is the spatial charge density, $Z_0$ is the vacuum impedance, $c$ is the light velocity in vacuum, and $\varepsilon_{\text{DC}}$ is the static dielectric function. $E_{\text{SC}}$ accelerates the photoexcited electrons to a local drift velocity $v_e(z) = -\mu_e E_{\text{SC}}(z) = z/\tau_{j_z}$, where $\mu_e > 0$ is the bulk electron mobility and

$$\tau_{j_z} = \frac{\varepsilon_{\text{DC}}}{Z_0 c e \mu_e n_0}. \tag{2}$$

The resulting current density $j_z(z,t) = \Delta(N_\uparrow + N_\downarrow)(t) v_e(z)$ and its dynamics are determined by a continuity equation and possible relaxation of the photoexcited carriers (Supplemental Section V). At the TI/air interface ($z = d$), we obtain $j_z(d,t) = A_{j_z} \Theta(t) e^{-t/\tau_{j_z}}$ with $A_{j_z} = \Delta(N_\uparrow + N_\downarrow)_0 d/\tau_{j_z}$ and the Heaviside step function $\Theta(t)$. Therefore, $\tau_{j_z}$ sets the relaxation time of the charge current $j_z$ and, likewise, spin current $j_s = j_z P_s$. It equals the time for the photoexcited electrons to drift toward the surface.

(A3) The $j_s$ arriving at the surface $z = d$ leads to spin accumulation in the TSS, which, due to spin-velocity locking, results in an in-plane charge current with sheet density $J_{x-}$ proportional to the accumulated spin sheet density [Fig. 4(c)]. Assuming the surface current flows at the TSS group velocity $v_{\text{TSS}}$ and decays with the time constant $\tau_{\text{TSS}}$, we find that

$$J_{x-}^{(A)}(t) = [H_{\text{TSS}} * (j_z P_s)](t)|_{z=d}. \tag{3}$$

Thus, $J_{x-}^{(A)}(t)$ equals the convolution ($*$) of the spin current $j_s = j_z P_s$ at $z = d$ and the TSS relaxation function $H_{\text{TSS}}(t) = v_{\text{TSS}} \Theta(t) e^{-t/\tau_{\text{TSS}}}$ [39]. From right to left, the three terms of Eq. (3) summarize the three steps discussed above: (A1) Spin polarization $P_s \propto B_{\text{ext}}$, (A2) drift of the photoexcited electrons in the space-charge field [$j_z$, Fig. 4(b)], and (A3) SCC in TSS [$H_{\text{TSS}}$, Fig. 4(c)].

We now compare the modeled $J_{x-}^{(A)}$ [Eq. (3)] to our experiment. If $\tau_{\text{TSS}} \ll \tau_{j_z}$, $J_{x-}^{(A)}$ rises and decays exponentially with the time constants $\tau_{\text{TSS}}$ and $\tau_{j_z}$, respectively. As seen in Fig. 3(c), we obtain excellent fits to our measured $J_{x-}$ waveforms, where $\tau_{\text{TSS}}$, $\tau_{j_z}$ and an overall amplitude scaling are the relevant fit parameters. The extracted $\tau_{\text{TSS}} \approx 100$ fs agrees with values inferred from previous THz-conductance measurements [24,40]. The modeled ($v_{\text{TSS}} \tau_{\text{TSS}} A_{j_z} P_s \approx 8$ A/m) and measured peak magnitude ($\approx 5$ A/m) of the current sheet density $J_{x-}$ agree well, too.

The relaxation time of the magneto-photocurrent $J_{x-}$ increases linearly with the TI thickness $d$ [inset of Fig. 3(c)]. This behavior is in line with $\tau_{j_z} \propto 1/n_0$ [see Eq. (2)] and the space-charge density $n_0 = N_{\text{SC}}/d$ (Fig. S10), i.e., the duration $\tau_{j_z}$ increases with $d$. From the slope $\tau_{j_z}/d \approx 10$ fs/nm of the measured $\tau_{j_z}$ vs $d$, we infer $N_{\text{SC}} \approx 1.3 \times 10^{12}$ cm$^{-2}$. This estimation is consistent with the Selenium vacancy density of $\sim 10^{12}$ cm$^{-2}$ obtained by scanning tunneling microscopy [27].

The decay $\tau_{j_z}$ of $J_{x-}$ extracted from (In$_r$Bi$_{1-r}$)$_2$Se$_3$(30 QL) depends only weakly on $r$ [Fig. S6(e)]. This observation suggests that the space-charge field remains persistent vs $r$. The photocurrent relaxation time of $\tau_{j_z} \lesssim 0.5$ ps is substantially faster than the typical electron-hole recombination time ($\approx 2$ ps) of Bi$_2$Se$_3$ [41]. Consequently, the latter is not visible in our current traces [Fig. 3(c)]. Interestingly, in the ultrafast ARPES signal from TSS [41], the extended rise time of $\approx 0.5$ ps could be interpreted as fingerprint of $\tau_{j_z}$ and, thus, the photocurrent $j_z$.

To summarize, we find good agreement between the observed and modeled dynamics, time constants and magnitude of the magneto-photocurrent sheet density $J_{x-}$. This agreement indicates that scenario (A) captures the crucial features of the underlying mechanisms.

**Scenario (B).** In this two-step scenario, the (B1) pump-induced out-of-plane drift charge current $j_z$ is (B2) converted into a transverse charge current by the ordinary Hall effect (ordHE) [Fig. 4(d)]. The conversion is characterized by the Hall angle $\gamma_{B_{\text{ext}}} \sim \mu_e B_{\text{ext}}$ and the momentum relaxation time $\tau_{\text{bulk}}$ of bulk electrons [33]. The resulting current sheet density is given by

$$J_{x-}^{(B)}(t) = \left(H_{\text{ordHE}} * \int \mathrm{d}z\, j_z\right)(t) \tag{4}$$

with $H_{\text{ordHE}}(t) = (\gamma_{B_{\text{ext}}}/\tau_{\text{bulk}})\Theta(t)e^{-t/\tau_{\text{bulk}}}$. The $z$ integration arises because the ordHE proceeds through the whole Bi$_2$Se$_3$ thickness. According to Eq. (4), $J_{x-}^{(B)}$ rises and decays exponentially with the time constants of, respectively, $\tau_{\text{bulk}}$ and $\tau_{j_z}$. This shape agrees with the experiment [Fig. 3(c)]. Further, the estimated peak $J_{x-}^{(B)}$ ($\approx 3$ A/m) is comparable with the measured peak current ($\approx 5$ A/m).

**Discussion.** The agreement of scenario (B) and our experimental data should be taken with caution. We first note that the modeled $J_{x-}^{(B)}$ is overestimated because our model ignores that electrons and holes contribute to $J_{x-}^{(B)}$ with opposite sign [42]. Second, the Drude relaxation time $\tau_{\text{bulk}} = 37$ fs of the Bi$_2$Se$_3$ bulk (Table S1) should set the rise time of $J_{x-}^{(B)}$ [Eq. (4)] [33]. In contrast, we observe that the rise time of $J_{x-}$ remains at a significantly larger value of $\approx 100$ fs for $r \leq 10\%$ of (In$_r$Bi$_{1-r}$)$_2$Se$_3$ [Fig. S6(d) and S6(e)].

Third, and most importantly, the in-plane charge currents in scenario (A) and (B) are generated, respectively, at the Bi$_2$Se$_3$ surface [Eq. (3)] and in the Bi$_2$Se$_3$ bulk [Eq. (4)]. Therefore, the photocurrent $J_{x-}$ scales differently with the sample thickness in the two scenarios. As seen in Fig. 3(b), scenario (A) explains the saturation of the photocurrent for $d > 30$ QL. In contrast, scenario (B) implies a quasilinear increase with $d$ (Supplemental Section IV). These three arguments indicate that scenario (B) makes a minor contribution to the observed magneto-photocurrent.

Scenario (A) is also consistent with the impact of Indium substitution (Fig. 2): As $r$ increases, both the amplitude of the THz signals $S_{x-}$ [Fig. 2(a-e)] and the TSS intensity in the ARPES spectra [Fig. 2(f-j)] decrease monotonically. Qualitatively, this correlation arises because the TSS contribute strongly to SCC at the TI surface. Quantitatively, it is highly nontrivial to translate the ARPES intensity of the TSS into a SCC strength. For example, for $r = 10\%$, the Dirac point is not visible any more [Fig. 2(i)], whereas $S_{x-}$ is still substantial [Fig. 2(d)]. This observation suggests the presence of spin-velocity locking without TSS but due to interfacial inversion asymmetry and strong spin-orbit coupling [43]. At $r = 20\%$, spin-orbit coupling has further decreased, resulting in a vanishing THz signal [Fig. 2(e)].

**In conclusion**, we observe THz magneto-photocurrents in the three-dimensional topological insulator Bi$_2$Se$_3$. They can consistently be explained by a spin-polarized out-of-plane current of optically excited electrons that is converted into an in-plane charge current by SCC at the TI surface. Interestingly, at a magnetic field 0.3 T, the magneto-photocurrent is only one order of magnitude smaller than in stacks of ferromagnetic Co and the TI Bi$_2$Te$_3$ [34], even though the out-of-plane spin current from a paramagnetic material is significantly smaller than from an optically heated ferromagnet. We believe that this relatively strong amplitude is a result of very efficient spin injection from the Bi$_2$Se$_3$ bulk into the TSS.

Importantly, from an applied viewpoint, THz magneto-photocurrents allow us to characterize the SCC efficiency of TSS without the need to attach ferromagnetic thin films that may strongly perturb the TI. We expect that our method can be transferred to other three-dimensional TIs to study their SCC capabilities. Even more generally, our methodology is interesting for measuring SCC in materials and sample systems with paramagnetic rather than spontaneous magnetic ordering. Thus, it strongly widens the scope of spintronic THz emission and may lead to the development of spintronic THz emitters without ferromagnets [44].

**Acknowledgments**

C.I. was supported by the National Research Foundation of Korea (Grant No. 2021R1A6A3A14044225). D.J. and S.O. was supported by the Army Research Office's W911NF2010108 and MURI W911NF2020166. G.B., T.S.S. and T.K. acknowledge funding by the European Research Council through ERC-2023 AdG ORBITERA (Grant No. 101142285), the European Union H2020 program through the FET project SKYTOP (Grant No. 824123) and by the Deutsche Forschungsgemeinschaft (DFG, German Research Foundation) through the Collaborative Research Center SFB TRR 227 "Ultrafast spin dynamics" (Project No. 328545488; projects A05 and B02) and the Excellence Cluster EXC 3112 "Center for Chiral Electronics" (EXC 3112/1, Project No. 533767171).

**References**

[1] Q. L. He, T. L. Hughes, N. P. Armitage, Y. Tokura, and K. L. Wang, *Topological Spintronics and Magnetoelectronics*, Nat. Mater. **21**, 15 (2022).

[2] H. Wang et al., *Room Temperature Energy-Efficient Spin-Orbit Torque Switching in Two-Dimensional van Der Waals $Fe_3GeTe_2$ Induced by Topological Insulators*, Nat. Commun. **14**, 5173 (2023).

[3] D. Chaudhuri, M. Salehi, S. Dasgupta, M. Mondal, J. Moon, D. Jain, S. Oh, and N. P. Armitage, *Ambipolar Magneto-Optical Response of Ultralow Carrier Density Topological Insulators*, Phys. Rev. B **103**, L081110 (2021).

[4] R. Fujimura, R. Yoshimi, M. Mogi, A. Tsukazaki, M. Kawamura, K. S. Takahashi, M. Kawasaki, and Y. Tokura, *Current-Induced Magnetization Switching at Charge-Transferred Interface between Topological Insulator $(Bi,Sb)_2Te_3$ and van Der Waals Ferromagnet $Fe_3GeTe_2$*, Appl. Phys. Lett. **119**, 032402 (2021).

[5] P. He, S. S. L. Zhang, D. Zhu, S. Shi, O. G. Heinonen, G. Vignale, and H. Yang, *Nonlinear Planar Hall Effect*, Phys. Rev. Lett. **123**, 16801 (2019).

[6] H. Soifer et al., *Band-Resolved Imaging of Photocurrent in a Topological Insulator*, Phys. Rev. Lett. **122**, 167401 (2019).

[7] K. N. Okada, N. Ogawa, R. Yoshimi, A. Tsukazaki, K. S. Takahashi, M. Kawasaki, and Y. Tokura, *Enhanced Photogalvanic Current in Topological Insulators via Fermi Energy Tuning*, Phys. Rev. B **93**, 081403(R) (2016).

[8] M. Jamali, J. S. Lee, J. S. Jeong, F. Mahfouzi, Y. Lv, Z. Zhao, B. K. Nikolić, K. A. Mkhoyan, N. Samarth, and J. P. Wang, *Giant Spin Pumping and Inverse Spin Hall Effect in the Presence of Surface and Bulk Spin-Orbit Coupling of Topological Insulator $Bi_2Se_3$*, Nano Lett. **15**, 7126 (2015).

[9] S. Zhang and A. Fert, *Conversion between Spin and Charge Currents with Topological Insulators*, Phys. Rev. B **94**, 184423 (2016).

[10] E. Longo, L. Locatelli, P. Tsipas, A. Lintzeris, A. Dimoulas, M. Fanciulli, M. Longo, and R. Mantovan, *Exploiting the Close-to-Dirac Point Shift of the Fermi Level in the $Sb_2Te_3/Bi_2Te_3$ Topological Insulator Heterostructure for Spin-Charge Conversion*, ACS Appl. Mater. Interfaces **15**, 50237 (2023).

[11] E. Longo et al., *Large Spin-to-Charge Conversion at Room Temperature in Extended Epitaxial $Sb_2Te_3$ Topological Insulator Chemically Grown on Silicon*, Adv. Funct. Mater. **32**, 2109361 (2022).

[12] A. R. Mellnik et al., *Spin-Transfer Torque Generated by a Topological Insulator*, Nature **511**, 449 (2014).

[13] Y. Wang et al., *Room Temperature Magnetization Switching in Topological Insulator-Ferromagnet Heterostructures by Spin-Orbit Torques*, Nat. Commun. **8**, 1364 (2017).

[14] H. Wu et al., *Room-Temperature Spin-Orbit Torque from Topological Surface States*, Phys. Rev. Lett. **123**, 207205 (2019).

[15] H. Wang, J. Kally, J. S. Lee, T. Liu, H. Chang, D. R. Hickey, K. A. Mkhoyan, M. Wu, A. Richardella, and N. Samarth, *Surface-State-Dominated Spin-Charge Current Conversion in Topological-Insulator-Ferromagnetic-Insulator Heterostructures*, Phys. Rev. Lett. **117**, 076601 (2016).

[16] X. Wang et al., *Ultrafast Spin-to-Charge Conversion at the Surface of Topological Insulator Thin Films*, Adv. Mater. **30**, 1802356 (2018).

[17] E. Rongione et al., *Ultrafast Spin-Charge Conversion at SnBi$_2$Te$_4$/Co Topological Insulator Interfaces Probed by Terahertz Emission Spectroscopy*, Adv. Opt. Mater. **10**, 2102061 (2022).

[18] H. Park et al., *Enhanced Spin-to-Charge Conversion Efficiency in Ultrathin Bi$_2$Se$_3$ Observed by Spintronic Terahertz Spectroscopy*, ACS Appl. Mater. Interfaces **13**, 23153 (2021).

[19] X. Chen et al., *Generation and Control of Terahertz Spin Currents in Topology-Induced 2D Ferromagnetic Fe$_3$GeTe$_2$|Bi$_2$Te$_3$ Heterostructures*, Adv. Mater. **34**, 2106172 (2022).

[20] A. Manchon, J. Železný, I. M. Miron, T. Jungwirth, J. Sinova, A. Thiaville, K. Garello, and P. Gambardella, *Current-Induced Spin-Orbit Torques in Ferromagnetic and Antiferromagnetic Systems*, Rev. Mod. Phys. **91**, 035004 (2019).

[21] A. Wolos et al., *G-Factors of Conduction Electrons and Holes in Bi$_2$Se$_3$ Three-Dimensional Topological Insulator*, Phys. Rev. B **93**, 155114 (2016).

[22] O. Ly and D. M. Basko, *Theory of Electron Spin Resonance in Bulk Topological Insulators Bi$_2$Se$_3$, Bi$_2$Te$_3$ and Sb$_2$Te$_3$*, J. Phys. Condens. Matter **28**, 155801 (2016).

[23] M. Brahlek, N. Bansal, N. Koirala, S. Y. Xu, M. Neupane, C. Liu, M. Z. Hasan, and S. Oh, *Topological-Metal to Band-Insulator Transition in (Bi$_{1-x}$In$_x$)$_2$Se$_3$ Thin Films*, Phys. Rev. Lett. **109**, 186403 (2012).

[24] L. Wu, M. Brahlek, R. Valdés Aguilar, A. V Stier, C. M. Morris, Y. Lubashevsky, L. S. Bilbro, N. Bansal, S. Oh, and N. P. Armitage, *A Sudden Collapse in the Transport Lifetime across the Topological Phase Transition in (Bi$_{1-x}$In$_x$)$_2$Se$_3$*, Nat. Phys. **9**, 410 (2013).

[25] C. Heide et al., *Probing Topological Phase Transitions Using High-Harmonic Generation*, Nat. Photonics **16**, 620 (2022).

[26] N. Bansal, Y. S. Kim, M. Brahlek, E. Edrey, and S. Oh, *Thickness-Independent Transport Channels in Topological Insulator Bi$_2$Se$_3$ Thin Films*, Phys. Rev. Lett. **109**, 116804 (2012).

[27] N. Koirala et al., *Record Surface State Mobility and Quantum Hall Effect in Topological Insulator Thin Films via Interface Engineering*, Nano Lett. **15**, 8245 (2015).

[28] Y. Cao et al., *Mapping the Orbital Wavefunction of the Surface States in Three-Dimensional Topological Insulators*, Nat. Phys. **9**, 499 (2013).

[29] T. S. Seifert et al., *Femtosecond Formation Dynamics of the Spin Seebeck Effect Revealed by Terahertz Spectroscopy*, Nat. Commun. **9**, 2899 (2018).

[30] L. Braun, G. Mussler, A. Hruban, M. Konczykowski, T. Schumann, M. Wolf, M. Münzenberg, L. Perfetti, and T. Kampfrath, *Ultrafast Photocurrents at the Surface of the Three-Dimensional Topological Insulator Bi$_2$Se$_3$*, Nat. Commun. **7**, 13259 (2016).

[31] K. Takahashi, T. Kanno, A. Sakai, H. Tamaki, H. Kusada, and Y. Yamada, *Terahertz Radiation via Ultrafast Manipulation of Thermoelectric Conversion in Thermoelectric Thin Films*, Adv. Opt. Mater. **2**, 428 (2014).

[32] T. Kampfrath et al., *Terahertz Spin Current Pulses Controlled by Magnetic Heterostructures*, Nat. Nanotechnol. **8**, 256 (2013).

[33] J. Shan, T. F. Heinz, C. Weiss, R. Wallenstein, and R. Beigang, *Origin of Magnetic Field Enhancement in the Generation of THz Radiation from Semiconductor Surfaces*, Opt. Lett. **26**, 849 (2001).

[34] G. Bierhance et al., *Terahertz Time-Domain Signatures of the Inverse Edelstein Effect in Topological Insulator|ferromagnet Heterostructures*, ArXiv:2506.22327 (2025).

[35] D. Hsieh et al., *Observation of Time-Reversal-Protected Single-Dirac-Cone Topological-Insulator States in $Bi_2Te_3$ and $Sb_2Te_3$*, Phys. Rev. Lett. **103**, 146401 (2009).

[36] D. Hsieh et al., *A Tunable Topological Insulator in the Spin Helical Dirac Transport Regime*, Nature **460**, 1101 (2009).

[37] J. W. McIver, D. Hsieh, S. G. Drapcho, D. H. Torchinsky, D. R. Gardner, Y. S. Lee, and N. Gedik, *Theoretical and Experimental Study of Second Harmonic Generation from the Surface of the Topological Insulator $Bi_2Se_3$*, Phys. Rev. B **86**, 035327 (2012).

[38] D. Hsieh, J. W. McIver, D. H. Torchinsky, D. R. Gardner, Y. S. Lee, and N. Gedik, *Nonlinear Optical Probe of Tunable Surface Electrons on a Topological Insulator*, Phys. Rev. Lett. **106**, 057401 (2011).

[39] J. Reimann et al., *Subcycle Observation of Lightwave-Driven Dirac Currents in a Topological Surface Band*, Nature **562**, 396 (2018).

[40] R. V. Aguilar et al., *Terahertz Response and Colossal Kerr Rotation from the Surface States of the Topological Insulator $Bi_2Se_3$*, Phys. Rev. Lett. **108**, 087403 (2012).

[41] Y. H. Wang, D. Hsieh, E. J. Sie, H. Steinberg, D. R. Gardner, Y. S. Lee, P. Jarillo-Herrero, and N. Gedik, *Measurement of Intrinsic Dirac Fermion Cooling on the Surface of the Topological Insulator $Bi_2Se_3$ Using Time-Resolved and Angle-Resolved Photoemission Spectroscopy*, Phys. Rev. Lett. **109**, 127401 (2012).

[42] L. G. Zhu, B. Kubera, K. Fai Mak, and J. Shan, *Effect of Surface States on Terahertz Emission from the $Bi_2Se_3$ Surface*, Sci. Rep. **5**, 10308 (2015).

[43] G. Bihlmayer, P. Noël, D. V. Vyalikh, E. V. Chulkov, and A. Manchon, *Rashba-like Physics in Condensed Matter*, Nat. Rev. Phys. **4**, 642 (2022).

[44] T. S. Seifert, L. Cheng, Z. Wei, T. Kampfrath, and J. Qi, *Spintronic Sources of Ultrashort Terahertz Electromagnetic Pulses*, Appl. Phys. Lett. **120**, 180401 (2022).

**Figures**

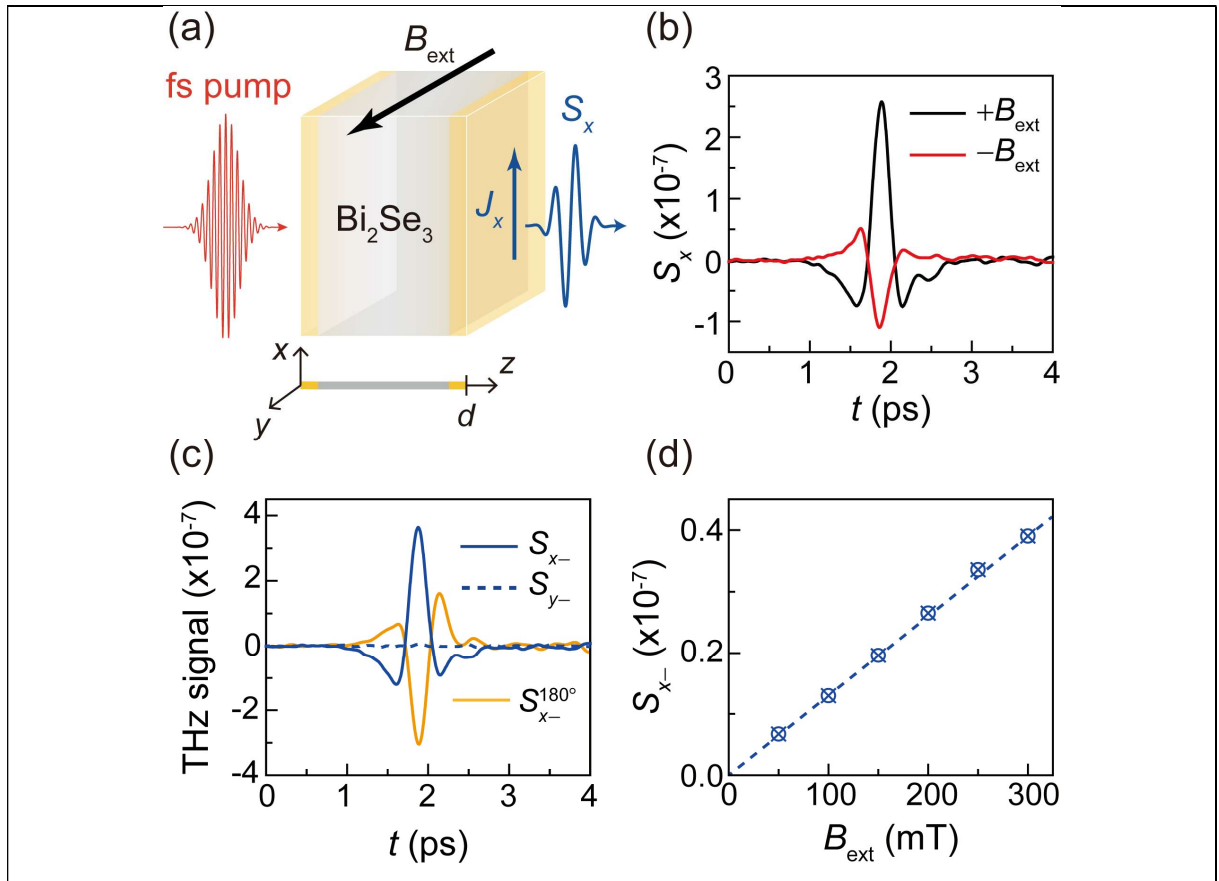

FIG. 1. (a) Schematic of the experiment. A femtosecond (fs) laser pulse drives an ultrafast photocurrent with density $\boldsymbol{j}(z,t)$ in a thin film of the topological insulator (TI) $Bi_2Se_3$(16 QL). By detecting the terahertz (THz) electric field emitted by $\boldsymbol{j}$, we can retrieve the dynamics of the current sheet density $\boldsymbol{J}(t) = \int dz\, \boldsymbol{j}(z,t)$ vs time $t$. An external magnetic field $\boldsymbol{B}_{\text{ext}} = B_{\text{ext}} \boldsymbol{u}_y$ with amplitude $B_{\text{ext}} = 0.3$ T is applied parallel to the $y$ axis. Gray and yellow regions indicate the $Bi_2Se_3$ bulk and the topologically protected surface states (TSS), respectively. (b) Typical electro-optic THz signals $S_x(t, \pm B_{\text{ext}})$ of the $x$-polarized THz electric-field component [see panel (a)]. (c) Waveforms of the THz electro-optic signals $S_{x-}(t)$ and $S_{y-}(t)$ odd in $B_{\text{ext}}$ [see Eq. (1)]. The signal $S_{x-}^{180°}(t)$ from the 180°-turned sample was corrected for the THz transmission through the substrate. (d) Root-mean-square amplitudes of $S_{x-}$ vs $B_{\text{ext}}$ along with a linear fit.

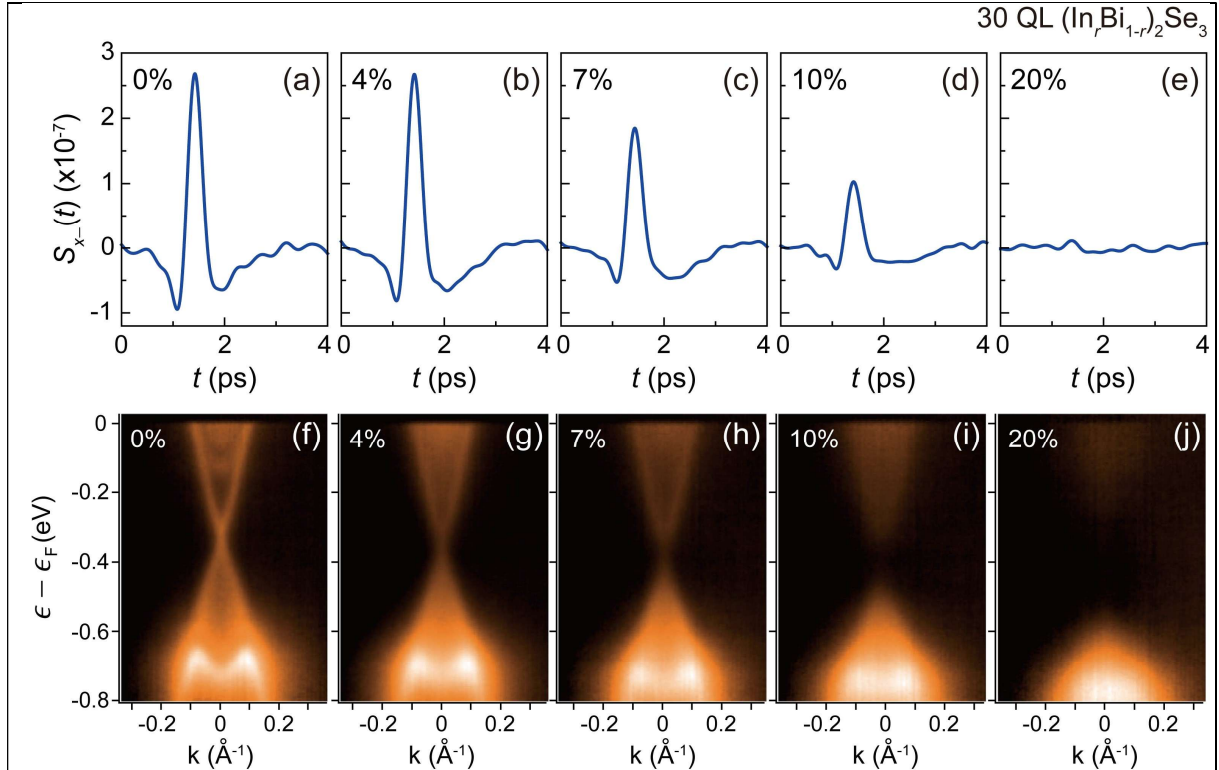

FIG. 2. Impact of Indium substitution. (a-e) THz magneto-photocurrent signal $S_{x-}(t)$ from $(In_rBi_{1-r})_2Se_3$(30 QL) with an Indium content of (a) $r = 0\%$, (b) 4%, (c) 7%, (d) 10% and (e) 20%. The external magnetic field has magnitude $|B_{ext}| = 0.3$ T. (f-j) Photoelectron emission intensity vs in-plane electron wavevector $k$ and energy $\epsilon - \epsilon_F$ from identical samples. The spectra were obtained in Ref. [25] for (f) $r = 0\%$, (g) 4%, (h) 7%, (i) 10% and (j) 20% without external magnetic field. They show the disappearance of the TSS at $r > 7\%$ [25]. Panels (f)-(j) are adapted with permission from Ref. [25], Springer Nature.

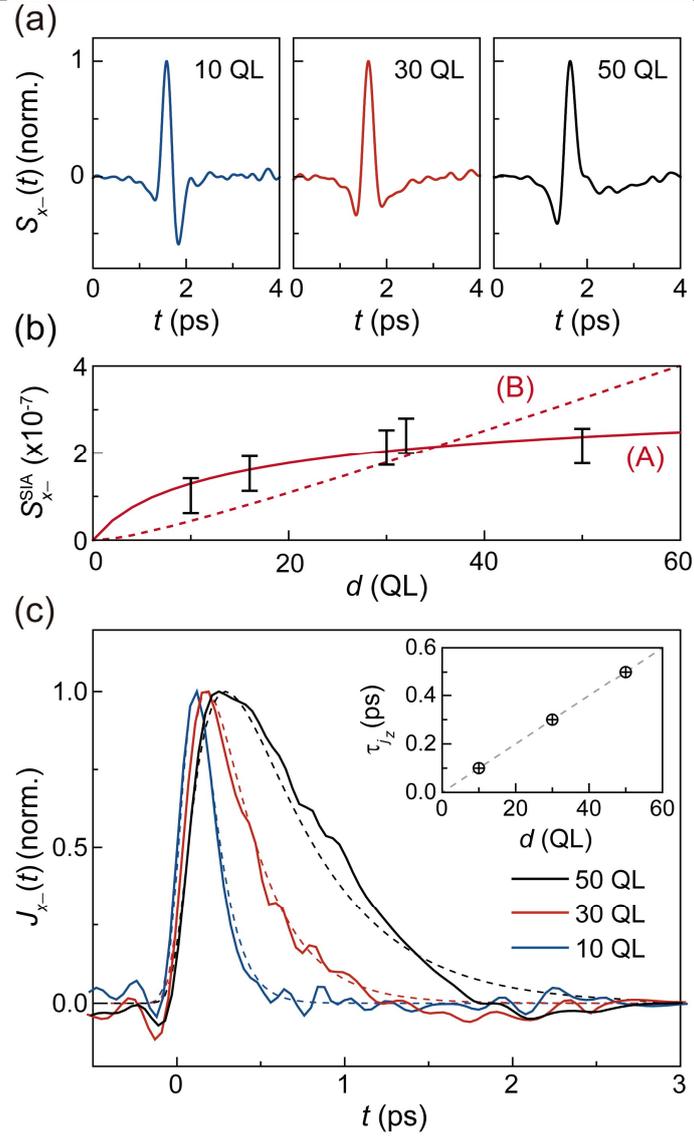

FIG. 3. Bi$_2$Se$_3$ thickness variation. (a) THz signals $S_{x-}(t)$ from Bi$_2$Se$_3(d)$ for $d = 10$, 30, and 50 QL. Signals are normalized to 1. (b) Amplitude of the THz signal $S_{x-}^{\mathrm{SIA}}$, i.e., the component of $S_{x-}$ that arises from the SIA of the sample. Fits based on model scenario (A) (solid red line) and scenario (B) (dashed red line) are also shown. (c) Current sheet density $J_{x-}(t)$ of the THz magneto-photocurrent extracted from $S_{x-}(t)$. Dashed lines are fits to the data (see main text). The inset shows the extracted relaxation time $\tau_{j_z}$ of the photocurrent vs $d$.

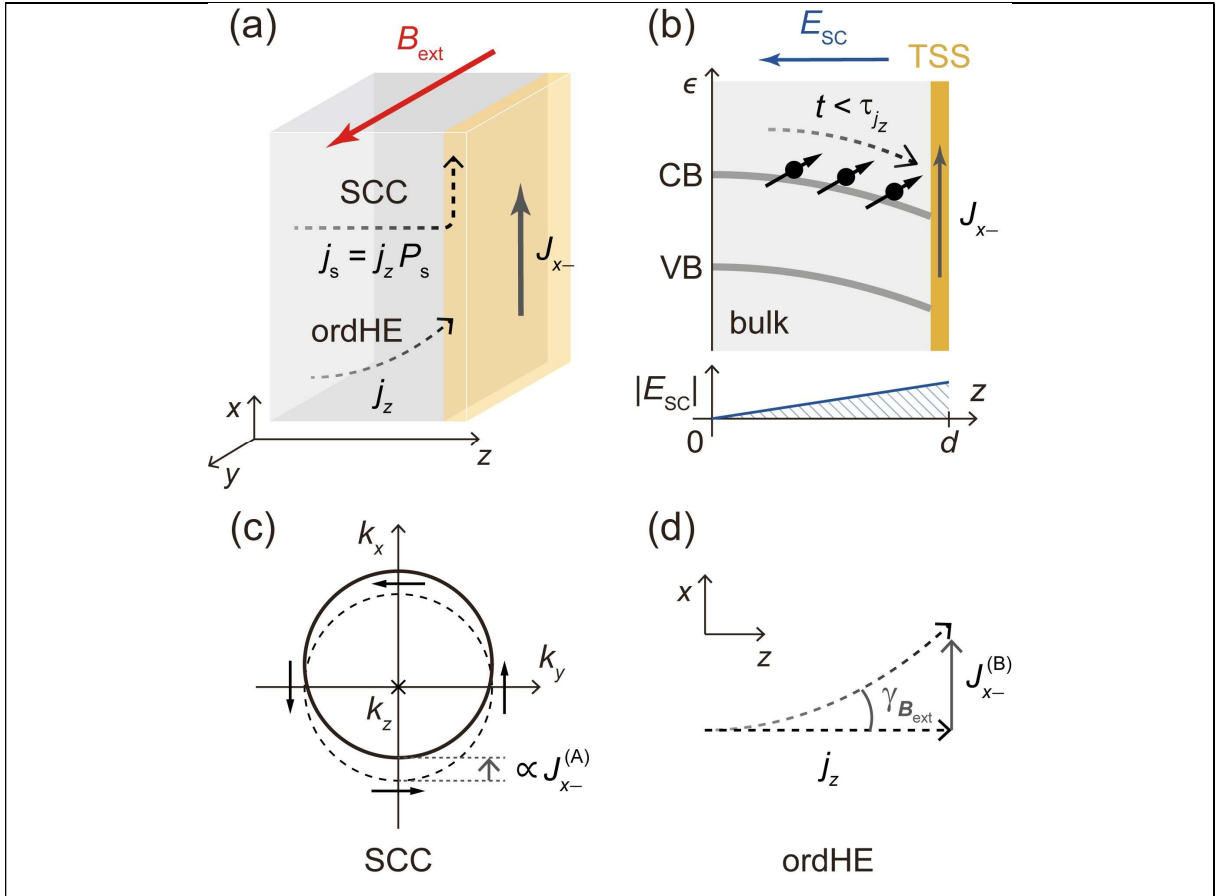

FIG. 4. Suggested scenarios (A) and (B) of the ultrafast magneto-photocurrent. (a) The pump pulse triggers a bulk charge current with density $j_z$ and a spin current with density $j_s = j_z P_s$ and $P_s \propto B_{ext}$. Both flow toward the Bi$_2$Se$_3$ surface. In (A), spin-charge conversion (SCC) in topological surface states (TSS) converts $j_s$ into a surface photocurrent with sheet density $J_{x-}^{(A)}$. In (B), the ordinary Hall effect (ordHE) due to $B_{ext}$ converts $j_z$ into a bulk current with sheet density $J_{x-}^{(B)} \propto B_{ext}$. (b) Qualitative $z$-dependence of the bottom of the bulk conduction band (CB) and the top of the valence band (VB) of Bi$_2$Se$_3$. The built-in electrostatic field $E_{SC}$ drives photoexcited electrons toward TSS at $z = d$ (yellow region) for the duration $\tau_{j_z}$. The TSS at the $z = 0$ interface are not shown. $|E_{SC}|$ increases with $z$. (c) The injection of $j_s$ into the TSS shifts the TI Fermi contour due to spin-velocity locking. The spin texture is left-handed relative to the $z$ axis for the upper Dirac cone. The photocurrent $J_{x-}^{(A)}$ flows for $\tau_{TSS} < t < \tau_{j_z}$, where $\tau_{TSS} \approx 100$ fs and $\tau_{j_z} \leq 0.5$ ps. (d) The ordinary Hall effect converts $j_z$ into $J_{x-}^{(B)}$ with a deflection angle $\gamma_{B_{ext}}$. The photocurrent $J_{x-}^{(B)}$ flows for $\tau_{bulk} < t < \tau_{j_z}$.